\documentclass[%
 reprint,
superscriptaddress,
 amsmath,amssymb,
 aps,
]{revtex4-2}

\usepackage{graphicx}% Include figure files
\usepackage{dcolumn}% Align table columns on decimal point
\usepackage{bm}% bold math
\usepackage{placeins}
\usepackage{orcidlink}
\usepackage{soul}
\usepackage{xcolor}

\begin{document}

\preprint{APS/123-QED}

\title{Symmetry-Based Design Rules for Second-Harmonic Generation in Stacked and Twisted MoS$_2$ Bilayers}% Force line breaks with \\


\author{Sumanti Patra\orcidlink{0000-0001-9178-4709}}
\email{sumanti1357@gmail.com}
\affiliation{Institute of Condensed Matter Theory and Optics,
Friedrich Schiller University Jena, 07743 Jena, Germany}
\author{Caterina Cocchi\orcidlink{0000-0002-9243-9461}}
\email{caterina.cocchi@uni-jena.de}
\affiliation{Institute of Condensed Matter Theory and Optics,
Friedrich Schiller University Jena, 07743 Jena, Germany}
\affiliation{Abbe Center of Photonics, Friedrich-Schiller-Universit\"at Jena, 07745, Jena, Germany}

%\author{Author 1}
%\affiliation{
% Second institution 
%}%
%\affiliation{
%\affiliation{%

\date{\today}% It is always \today, today,

\begin{abstract}
Understanding how stacking controls the nonlinear optical response of two-dimensional materials is key to designing van der Waals heterostructures with tailored functionalities. Here, we establish a comprehensive symmetry-based framework mapping the structural configuration of MoS$_2$ bilayers across four point groups ($D_{3h}$, $D_{3d}$, $C_{3v}$, $C_3$) to their second-order susceptibility tensor $\chi^{(2)}$. Using group-theory arguments benchmarked against first-principles response-function calculations, we demonstrate how symmetry breaking controls the activation and suppression of individual tensor elements in these systems. We show that the emergence of the in-plane component $\chi_{xxx}$ in twisted configurations ($C_3$ group) induces a rigid azimuthal rotation of the second-harmonic generation polar lobes, which remains frequency-independent across the entire optical spectrum, locking to half of the structural twist angle. Our findings establish a direct, wavelength-independent optical route for twist-angle determination and provide a clear roadmap for engineering nonlinear optical responses in two-dimensional materials.

\end{abstract}

\maketitle

%\tableofcontents

\section{\label{sec:level1} Introduction}

Two-dimensional transition metal dichalcogenides (TMDs) have emerged as an appealing platform for nonlinear optics, combining broken inversion symmetry, strong light-matter interaction, and a lattice highly sensitive to specific stacking configurations ~\cite{li2013,autere2018,klimmer2021all,dogadov2022,herrmann2025nonlinear}. In their monolayer form, semiconducting TMDs with chemical formula MX$_2$ (with M = Mo, W; X = S, Se, Te) preferentially crystallize in the non-centrosymmetric point group $D_{3h}$~\cite{wilson1969,chhowalla2013,manzeli2017}. This inherent absence of inversion symmetry permits a non-vanishing second-order susceptibility tensor $\chi^{(2)}$ and enables second-harmonic generation (SHG) \cite{li2013,kumar2013,malard2013,zhang2020second,du2024nonlinear,huang2024second}. This technique represents a powerful, non-destructive probe for mapping layer counts, stacking order, and crystallographic orientations across a wide variety of layered materials and moiré superlattices~\cite{hsu2014,psilodimitrakopoulos2019,wang2019second,zhou2020nonlinear,paradisanos2021}. 

A bilayer stack is characterized by an additional, highly tunable structural degree of freedom: relative translation and rotation of the constituent monolayers, which dramatically expand the accessible symmetry landscape. Depending on the specific configuration, the resulting point group can range from highly symmetric $D_{3h}$ and centrosymmetric $D_{3d}$, to lower-symmetry $C_{3v}$ and $C_3$ twisted configurations ~\cite{li2013,kumar2013,hsu2014,paradisanos2022}. Since the analytical structure of the $\chi^{(2)}$ tensor is entirely determined by the crystal point group via Neumann's principle~\cite{boyd2008}, this stacking-controlled symmetry hierarchy translates into a systematic modulation of the nonlinear optical responses, ranging from complete suppression of SHG to the activation of new tensor components that are silent in the freestanding monolayer~\cite{li2013,kumar2013,hsu2014,shi20173r}. 

First-principles calculations of nonlinear susceptibilities have successfully captured the response of idealized TMDs~\cite{attaccalite2013,attaccalite2019,pike-patcher2021,ruan2024exciton}. However, full-fledged \textit{ab initio} evaluations of $\chi^{(2)}$ spectra become exceptionally costly or even unfeasible for realistic low-symmetry configurations and large-scale moiré superlattices ~\cite{gruning2014,wang2017wannier,garcia2023}. This computational bottleneck makes a rigorous symmetry-based analysis vital to map the nonlinear response of these systems across arbitrary structural settings without relying on brute-force calculations. Establishing such a framework on a qualitatively reliable and physically accessible baseline, as offered by the independent-particle approximation, provides the necessary foundation to subsequently understand the role of many-body effects and to develop numerically robust high-throughput screening workflows for SHG in realistic 2D material architectures.

To bridge this gap, we present a systematic study coupling this predictive symmetry analysis with targeted first-principles validations of the frequency-dependent angular response. Using MoS$_2$ as a representative member of the TMD family, we map monolayer and bilayer configurations across four distinct point groups ($D_{3h}$, $D_{3d}$, $C_{3v}$, and $C_3$) to their respective $\chi^{(2)}$ tensors. While polarization-resolved SHG rotation has previously been used to extract twist angles in TMD bilayers~\cite{paradisanos2022}, typically interpreted via phenomenological interference between independent monolayers, we apply Neumann's principle to derive the complete $\chi^{(2)}$ tensor structure by treating the stacked or twisted bilayer as a single composite crystal. This symmetry-first approach holds independently of interlayer coupling strength and unifies all translational and rotational stacking configurations within a single, self-consistent framework. Starting from pure group-theory arguments, we determine the allowed and forbidden tensor components for each symmetry class and define how these constraints dictate macroscopic SHG angular patterns. Benchmarking these analytical relations against first-principles calculations establishes a clear, transferable set of design rules connecting stacking configurations to nonlinear optical signatures, including complete SHG suppression in centrosymmetric configurations and the emergence of a spectrally invariant and geometrically locked SHG lobe shift in twisted bilayers.

%\begin{figure}[h]
\begin{figure*}
\includegraphics[width=\textwidth]{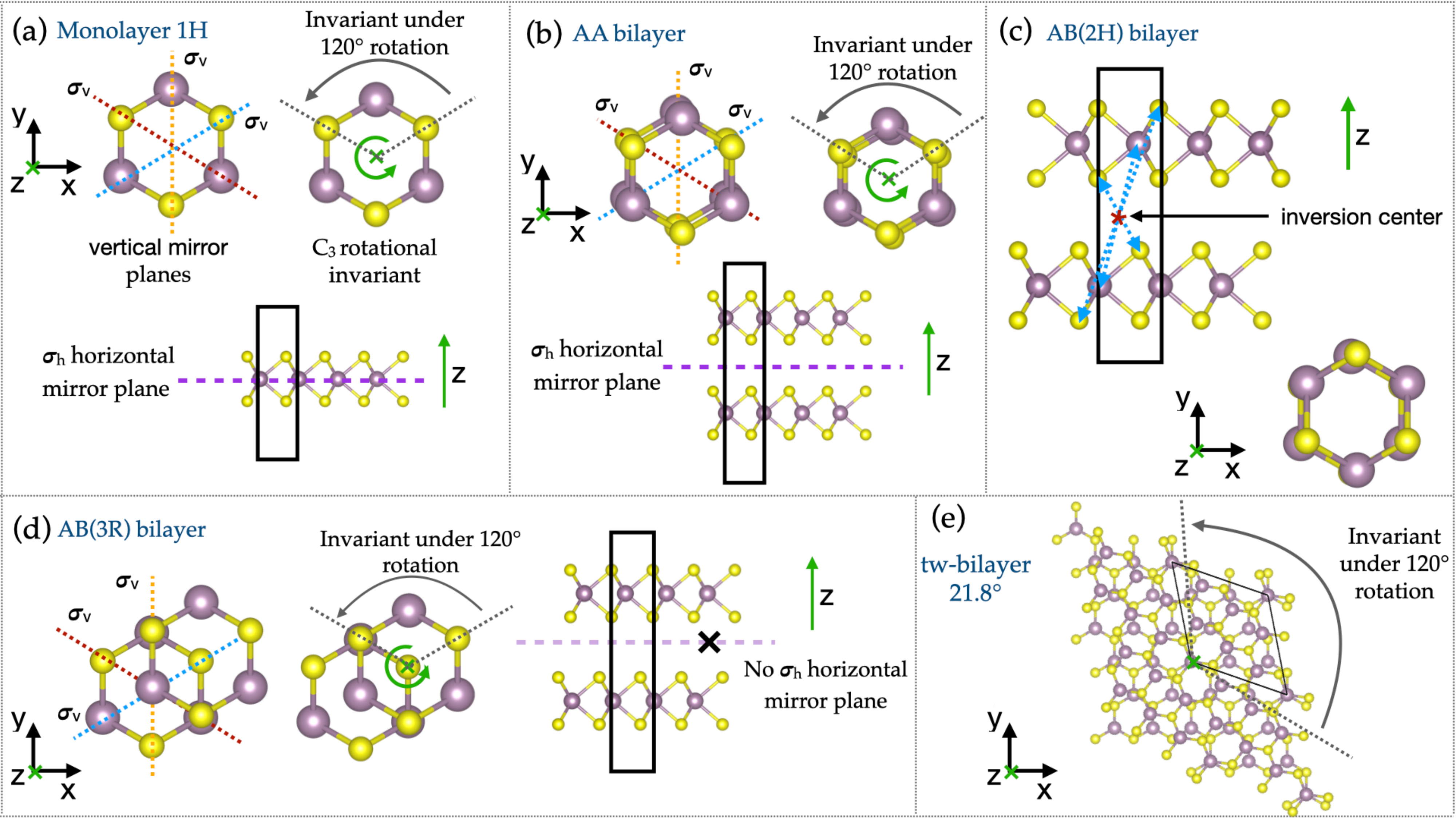}
\caption{Overview of the atomic structures and symmetry elements of the five MoS$_2$ configurations considered in this work. Mo and S atoms are shown in violet and yellow, respectively. (a) Monolayer 1H ($D_{3h}$): top view displaying the three vertical mirror planes $\sigma_v$ (colored dashed lines) alongside the $C_3$ rotational invariance, and side view showing the horizontal mirror plane $\sigma_h$ (purple dashed line). (b) Bilayer in the AA-stacking ($D_{3h}$), retaining all symmetry elements of the monolayer, including $\sigma_v$, $C_3$, and $\sigma_h$. (c) AB(2H)-stacked bilayer ($D_{3d}$), enabling the activation of an inversion center ($\mathcal{I}$, red asterisk). (d) AB(3R) bilayer ($C_{3v}$): $\sigma_v$ and $C_3$ symmetries are retained while the horizontal mirror plane $\sigma_h$ is broken.}
\label{fig:structures}
\end{figure*}

\section{\label{sec:level2}Symmetry Analysis and Design Rules}

\subsection{Crystal Structures and Point Groups}
\label{sec:design}

Monolayer MoS$_2$ in the 1H structure consists of a Mo atomic plane sandwiched between two S planes in a trigonal prismatic coordination. The resulting crystal belongs to the $D_{3h}$ point group, with symmetry elements including a threefold rotation axis $C_3$, a horizontal mirror plane $\sigma_h$, and three vertical mirror planes $\sigma_v$. Crucially, the lack of an inversion center permits a non-vanishing $\chi^{(2)}$ (Fig.~\ref{fig:structures}a). Forming a bilayer introduces an additional structural degree of freedom via the relative translation and rotation of the two layers, giving rise to a family of distinct stackings with qualitatively different symmetries. In the following analysis, we consider a selection of representative configurations (Fig.~\ref{fig:structures}b-e), belonging to different symmetry groups (Table~\ref{tab:pointgroups}). 

The AA--stacked bilayer, in which the top and bottom sheets are directly eclipsed, retains the same $D_{3h}$ symmetry of the parent monolayer (Fig.~\ref{fig:structures}b). In contrast, the AB(2H) stacking, associated with the thermodynamically stable form of bulk MoS$_2$ \cite{wilson1969,mak2010}, places the Mo atom of the top layer directly above the S atom of the bottom layer. This arrangement introduces an inversion center ($\mathcal{I}$) at the interlayer midpoint, elevating the point group to $D_{3d}$ and rendering the structure centrosymmetric (Fig.~\ref{fig:structures}c). Translating the layers into the AB(3R) configuration breaks both the inversion symmetry of the 2H-stacking and the horizontal reflection symmetry of the AA bilayer phase, yielding a non-centrosymmetric $C_{3v}$ point group where only the threefold rotation axis and the three vertical mirrors ($\sigma_v$) are preserved (Fig.~\ref{fig:structures}d). Finally, introducing a relative rotation between the layers destroys all mirror symmetries, $\sigma_h$ and $\sigma_v$. For instance, a twisted (tw) bilayer at the commensurate angle $\theta_{tw} = 21.8^\circ$ forms a coincidence site lattice supercell with 42 atoms, that retains only the threefold rotation axis (Fig.~\ref{fig:structures}e), reducing the symmetry to the $C_3$ point group (Table~\ref{tab:pointgroups}). 

The symmetry reduction from the highest-symmetry parent bilayers to the twisted configuration follows two separate pathways: 
\begin{align}
    %D_{3h} \supset C_{3v} \supset C_3
    & D_{3h} \xrightarrow{\text{translation}} C_{3v} \xrightarrow{\text{twisting}} C_3 \\
    %\qquad \text{and} \qquad
    %D_{3d} \supset C_{3v} \supset C_3,
    & D_{3d} \xrightarrow{\text{twisting}} C_3.
\end{align}
While relative interlayer translation breaks horizontal symmetries, rotational twisting eliminates all mirror planes and inversion centers, leaving the $C_3$ subgroup as the lowest common structural denominator for all twisted and moiré configurations.

\begin{table}%[h!]
\caption{Point groups and key symmetry elements for the targeted monolayer and bilayer MoS$_2$ configurations.}
\label{tab:pointgroups}
\begin{ruledtabular}
\begin{tabular}{llccccc}
System & Stacking & Point group & $C_3$ & $\sigma_h$ & $\sigma_v$ & 
$\mathcal{I}$ \\
\hline
Monolayer  & 1H          & $D_{3h}$ & \checkmark & \checkmark & \checkmark & $\times$ \\
Bilayer    & AA          & $D_{3h}$ & \checkmark & \checkmark & \checkmark & $\times$ \\
Bilayer    & AB(2H)          & $D_{3d}$ & \checkmark & $\times$   & \checkmark & \checkmark \\
Bilayer    & AB(3R)          & $C_{3v}$ & \checkmark & $\times$   & \checkmark & $\times$ \\
Twisted Bilayer & $\theta_{tw} = 21.8^\circ$ & $C_3$  & \checkmark & $\times$   & $\times$   & $\times$ \\
%Twisted BL & $38.2^\circ$ & $C_3$  & \checkmark & $\times$   & $\times$   & $\times$ \\
\end{tabular}
\end{ruledtabular}
\end{table}

\subsection{$\chi^{(2)}$ Tensor Structure from Neumann's Principle}

The second-order susceptibility $\chi^{(2)}_{ijk}(\omega)$ is a third-rank polar tensor relating the second-harmonic polarization to the external electric field:
\begin{equation}
    P_i(2\omega) =  \sum_{jk} \chi^{(2)}_{ijk}(\omega)\, E_j(\omega) E_k(\omega).
\end{equation}
Neumann's principle imposes the invariance of the tensor under any spatial symmetry operation $g$ belonging to the crystal point group~\cite{boyd2008}:
\begin{equation}
    \chi^{(2)}_{ijk} = R_{il}R_{jm}R_{kn}\,\chi^{(2)}_{lmn},
\end{equation}
where $R$ is the matrix representation of $g$ and summation over repeated indices is implied. Applying this constraint to the generators of each point group, while invoking the intrinsic permutation symmetry $\chi^{(2)}_{ijk} = \chi^{(2)}_{ikj}$ imposed by the indistinguishability of the two photons in the SHG process, yields independent, non-zero tensor components for each configuration.

Before evaluating individual point groups, we note that the presence of a threefold rotation axis ($C_3 \parallel z$) common to all five systems imposes a strong restriction to the 27 tensor components. Specifically, it forces all elements with in-plane index ($x$ or $y$) combined with two out-of-plane ($z$) indices to vanish identically (see details in the Appendix):
\begin{equation}
    \chi^{(2)}_{xzz} = \chi^{(2)}_{yzz} = \chi^{(2)}_{zxz} = \chi^{(2)}_{zzx} = 
    \chi^{(2)}_{zyz} = \chi^{(2)}_{zzy} = \chi^{(2)}_{zxy} = \chi^{(2)}_{zyx} = 0.
\end{equation}
Furthermore, components featuring mixed in-plane indices combined with a single $z$ index at the primary position, namely $\chi^{(2)}_{zxy}$ and $\chi^{(2)}_{zyx}$, are also strictly forbidden by $C_3$ symmetry. The remaining non-zero elements are linked via rotational invariants, which are subsequently lifted or constrained by the mirror planes and inversion centers of each point group (Table~\ref{tab:full27}).

%\FloatBarrier
\begin{table*}[t]
\caption{Classification of all 27 components of the $\chi^{(2)}_{ijk}$ tensor for each point group. The checkmark (\checkmark) denotes an independent, non-zero component. Formulations explicitly show the relationship dictated by the responsible symmetry element. $\times$($C_3$) indicates components forced to zero by the threefold rotation axis alone across all groups; $\times$($\sigma_h$) marks suppression via the horizontal mirror plane (odd number of $z$ indices). $\times$($\sigma_v$) indicates vertical mirror suppression. $\times$($\mathcal{I}$) indicates centrosymmetric elimination via inversion; ``perm'' denotes relations via intrinsic SHG permutation symmetry $\chi^{(2)}_{ijk}=\chi^{(2)}_{ikj}$.}
\label{tab:full27}
\begin{ruledtabular}
\begin{tabular}{lcccc}
Component & $D_{3h}$ & $D_{3d}$ & $C_{3v}$ & $C_3$ \\
\hline
\multicolumn{5}{l}{\textit{In-plane, $\sigma_v$-sensitive components}} \\
$\chi^{(2)}_{xxx}$ & $\times$($\sigma_v$) & $\times$($\mathcal{I}$) & $\times$($\sigma_v$) & \checkmark \\
$\chi^{(2)}_{xyy}$ & $\times$($\sigma_v$) & $\times$($\mathcal{I}$) & $\times$($\sigma_v$) & $=-\chi^{(2)}_{xxx}$($C_3$) \\
$\chi^{(2)}_{yxy}$ & $\times$($\sigma_v$) & $\times$($\mathcal{I}$) & $\times$($\sigma_v$) & $=-\chi^{(2)}_{xxx}$($C_3$) \\
$\chi^{(2)}_{yyx}$ & $\times$($\sigma_v$) & $\times$($\mathcal{I}$) & $\times$($\sigma_v$) & $=-\chi^{(2)}_{xxx}$($C_3$) \\
\hline
\multicolumn{5}{l}{\textit{In-plane, $C_3$-related components}} \\
$\chi^{(2)}_{yyy}$ & \checkmark & $\times$($\mathcal{I}$) & \checkmark & \checkmark \\
$\chi^{(2)}_{yxx}$ & $=-\chi^{(2)}_{yyy}$($C_3$) & $\times$($\mathcal{I}$) & $=-\chi^{(2)}_{yyy}$($C_3$) & $=-\chi^{(2)}_{yyy}$($C_3$) \\
$\chi^{(2)}_{xxy}$ & $=-\chi^{(2)}_{yyy}$($C_3$) & $\times$($\mathcal{I}$) & $=-\chi^{(2)}_{yyy}$($C_3$) & $=-\chi^{(2)}_{yyy}$($C_3$) \\
$\chi^{(2)}_{xyx}$ & $=-\chi^{(2)}_{yyy}$($C_3$) & $\times$($\mathcal{I}$) & $=-\chi^{(2)}_{yyy}$($C_3$) & $=-\chi^{(2)}_{yyy}$($C_3$) \\
\hline
\multicolumn{5}{l}{\textit{Out-of-plane, one $z$ index ($\sigma_h$-sensitive)}} \\
$\chi^{(2)}_{xxz}$ & $\times$($\sigma_h$) & $\times$($\mathcal{I}$) & \checkmark & \checkmark \\
$\chi^{(2)}_{xzx}$ & $\times$($\sigma_h$) & $\times$($\mathcal{I}$) & $=\chi^{(2)}_{xxz}$(perm) & $=\chi^{(2)}_{xxz}$(perm) \\
$\chi^{(2)}_{yyz}$ & $\times$($\sigma_h$) & $\times$($\mathcal{I}$) & $=\chi^{(2)}_{xxz}$($C_3$) & $=\chi^{(2)}_{xxz}$($C_3$) \\
$\chi^{(2)}_{yzy}$ & $\times$($\sigma_h$) & $\times$($\mathcal{I}$) & $=\chi^{(2)}_{xxz}$(perm+$C_3$) & $=\chi^{(2)}_{xxz}$(perm+$C_3$) \\
$\chi^{(2)}_{zxx}$ & $\times$($\sigma_h$) & $\times$($\mathcal{I}$) & \checkmark & \checkmark \\
$\chi^{(2)}_{zyy}$ & $\times$($\sigma_h$) & $\times$($\mathcal{I}$) & $=\chi^{(2)}_{zxx}$($C_3$) & $=\chi^{(2)}_{zxx}$($C_3$) \\
$\chi^{(2)}_{zzz}$ & $\times$($\sigma_h$) & $\times$($\mathcal{I}$) & \checkmark & \checkmark \\
\hline
\multicolumn{5}{l}{\textit{Mixed, one $z$ index, $\sigma_v$-sensitive}} \\
$\chi^{(2)}_{xyz}$ & $\times$($\sigma_v$) & $\times$($\mathcal{I}$) & $\times$($\sigma_v$) & \checkmark \\
$\chi^{(2)}_{xzy}$ & $\times$($\sigma_v$) & $\times$($\mathcal{I}$) & $\times$($\sigma_v$) & $=\chi^{(2)}_{xyz}$(perm) \\
$\chi^{(2)}_{yxz}$ & $\times$($\sigma_v$) & $\times$($\mathcal{I}$) & $\times$($\sigma_v$) & $=-\chi^{(2)}_{xyz}$($C_3$) \\
$\chi^{(2)}_{yzx}$ & $\times$($\sigma_v$) & $\times$($\mathcal{I}$) & $\times$($\sigma_v$) & $=-\chi^{(2)}_{xyz}$(perm+$C_3$) \\
\hline
\multicolumn{5}{l}{\textit{Zero by $C_3$ for all point groups under consideration.}} \\
$\chi^{(2)}_{xzz}$,$\chi^{(2)}_{yzz}$, $\chi^{(2)}_{zxz}$, $\chi^{(2)}_{zzx}$, $\chi^{(2)}_{zyz}$ , $\chi^{(2)}_{zzy}$, $\chi^{(2)}_{zxy}$, $\chi^{(2)}_{zyx}$ & $\times$($C_3$) & $\times$($C_3$) & $\times$($C_3$) & $\times$($C_3$) \\
\end{tabular}
\end{ruledtabular}
\end{table*}
%\FloatBarrier

\textit{$D_{3h}$ (monolayer, AA bilayer).} The horizontal mirror plane $\sigma_h$ maps coordinates as $z \to -z$, forcing all tensor elements containing an odd number of $z$ indices to vanish identically. The remaining purely in-plane components are simultaneously bounded by the $C_3$ axis and the vertical mirrors $\sigma_v$, reducing to:
\begin{equation}
    \chi^{(2)}_{yyy} = -\chi^{(2)}_{yxx} = -\chi^{(2)}_{xxy} = -\chi^{(2)}_{xyx}.
    \label{eq:chi2_D3h}
\end{equation}
All out-of-plane elements (e.g., $\chi^{(2)}_{zzz}$, $\chi^{(2)}_{zxx}$, $\chi^{(2)}_{xxz}$, and so on) vanish under $\sigma_h$, leaving the non-zero components in Eq.~\eqref{eq:chi2_D3h} to rule the entire nonlinear optical response of the system.

\textit{$D_{3d}$ (AB(2H) bilayer).} The presence of an inversion center $\mathcal{I}$ maps all coordinates as $x_i \to -x_i$, transforming any third-rank polar tensor component to its negative:
\begin{equation}
    \chi^{(2)}_{ijk} \xrightarrow{\mathcal{I}} -\chi^{(2)}_{ijk}.
\end{equation}
Combined with the invariance requirement, this yields:
\begin{equation}
    \chi^{(2)}_{ijk} = 0 \qquad \forall\; i,j,k.
    \label{eq:chi2_D3d}
\end{equation}
Hence, SHG is completely forbidden within the electric-dipole approximation for the AB(2H) stacking.

\textit{$C_{3v}$ (AB(3R) bilayer).} The absence of a $\sigma_h$ mirror plane activates tensor components with an odd number of $z$ indices. The three vertical mirrors $\sigma_v$ are retained, imposing $\chi^{(2)}_{xxx}=0$ and constraining the remaining in-plane elements to the same form as in the $D_{3h}$ group. Accounting for intrinsic permutation rules, the independent non-zero components are:
\begin{align}
    &\chi^{(2)}_{yyy} = -\chi^{(2)}_{yxx} = -\chi^{(2)}_{xxy} = -\chi^{(2)}_{xyx}, \nonumber \\
    &\chi^{(2)}_{zzz}, \nonumber \\
    &\chi^{(2)}_{zxx} = \chi^{(2)}_{zyy}, \nonumber \\
    &\chi^{(2)}_{xxz} = \chi^{(2)}_{xzx} = \chi^{(2)}_{yyz} = \chi^{(2)}_{yzy} .
    \label{eq:chi2_C3v}
\end{align}
The activation of out-of-plane elements ($\chi^{(2)}_{zzz}$ and $\chi^{(2)}_{zxx}$) makes the nonlinear response accessible via oblique-incidence or cross-polarized experimental setups, while the in-plane component matches the monolayer baseline.

\textit{$C_3$ (twisted bilayers).} In twisted configurations, only the threefold rotation axis survives, and the absence of any mirror symmetry operation distinguishes this case from all higher-symmetry stackings. The independent non-zero tensor components are:
\begin{align}
    &\chi^{(2)}_{yyy} = -\chi^{(2)}_{yxx} = -\chi^{(2)}_{xxy} = -\chi^{(2)}_{xyx}, 
    \nonumber \\
    &\chi^{(2)}_{xxx} = -\chi^{(2)}_{xyy} = -\chi^{(2)}_{yxy} = -\chi^{(2)}_{yyx}, 
    \nonumber \\
    &\chi^{(2)}_{zzz}, \nonumber \\
    &\chi^{(2)}_{zxx} = \chi^{(2)}_{zyy}, \nonumber \\
    &\chi^{(2)}_{xxz} = \chi^{(2)}_{yyz}, \nonumber \\
    &\chi^{(2)}_{xyz} = -\chi^{(2)}_{yxz}.
    \label{eq:chi2_C3}
\end{align}
The defining feature here is the simultaneous activation of $\chi^{(2)}_{xxx}$ alongside $\chi^{(2)}_{yyy}$. Since $\chi^{(2)}_{xxx} = 0$ is no longer enforced by $\sigma_v$, its emergence provides a direct optical signature of the structural twist. Additionally, the chiral out-of-plane components  ($\chi^{(2)}_{xyz}= -\chi^{(2)}_{yxz}$) become allowed, offering an additional degree of freedom unique to twisted moiré superlattices.

\subsection{Angular SHG Patterns and Lobe Shift}

For a normally incident fundamental beam with linear polarization at angle $\theta$ measured from the zigzag ($x$) axis, the co-polarized SHG intensity $I_\parallel(\theta)$ is determined by contracting the in-plane components of $\chi^{(2)}$ with the polarization vector $\hat{e} = (\cos\theta, \sin\theta, 0)$. 
The resulting second-harmonic polarization components are given by:
\begin{align}
    P_x(2\omega) &\propto \chi^{(2)}_{xxx}E_x^2 + \chi^{(2)}_{xyy}E_y^2 + 2\chi^{(2)}_{xxy}E_x E_y, \nonumber \\
    P_y(2\omega) &\propto \chi^{(2)}_{yxx}E_x^2 + \chi^{(2)}_{yyy}E_y^2 + 2\chi^{(2)}_{yxy}E_x E_y.
\end{align}
The co-polarized component is defined as $P_\parallel = P_x\cos\theta + P_y\sin\theta$, where the intensity scales as $I_\parallel \propto |P_\parallel|^2$.

\textit{$D_{3h}$ (monolayer, AA stacked bilayer).} Substituting the single independent component $\chi^{(2)}_{yyy}$ along with the symmetry-enforced relations $\chi^{(2)}_{xxy} = \chi^{(2)}_{xyx} = -\chi^{(2)}_{yyy}$ and $\chi^{(2)}_{yxx} = -\chi^{(2)}_{yyy}$ (Eq.~\ref{eq:chi2_D3h}) into the polarization equations yields:
\begin{align}
    P_x(2\omega) &\propto -2\chi^{(2)}_{yyy}E^2\cos\theta\sin\theta, \nonumber \\
    P_y(2\omega) &\propto \chi^{(2)}_{yyy}E^2(\sin^2\theta - \cos^2\theta).
        \label{eq:D3h_pxpy}
\end{align}
Projecting these components along the fundamental polarization axis gives:
\begin{equation}
\label{eq:p_parallel}
    P_\parallel \propto \chi^{(2)}_{yyy}\sin\theta\left[\sin^2\theta - 
    3\cos^2\theta\right] = -\chi^{(2)}_{yyy}\sin(3\theta),
\end{equation}
which yields a co-polarized SHG intensity profile of:
\begin{equation}
    I_\parallel(\theta) \propto |\chi^{(2)}_{yyy}|^2\sin^2(3\theta).
    \label{eq:D3h_pattern}
\end{equation}
This expression leads to the well-known six-lobed polar pattern with nodes spaced every $60^\circ$\cite{li2013,kumar2013,malard2013}. 
By aligning the $x$-axis with the zigzag direction, all intensity nodes fall exactly on the zigzag crystallographic axes ($\theta = 0^\circ, 60^\circ, 120^\circ, \dots$), whereas the maxima align with the armchair directions ($\theta = 30^\circ,90^\circ, 150^\circ, \dots$). This orientation provides direct crystallographic edge selectivity: a zigzag-terminated edge produces no co-polarized SHG signal, while an armchair edge yields the maximum possible emission.

\textit{$D_{3d}$ (2H bilayer).} Incorporating the condition $\chi^{(2)}_{ijk} = 0$ (Eq.~\ref{eq:chi2_D3d}) leads to:
\begin{equation}
    I_\parallel(\theta) = 0 \qquad \forall \;\theta,
\end{equation}
representing an identically vanishing polar response across all angles due to destructive interlayer interference.

\textit{$C_{3v}$ (3R bilayer).} Since the in-plane tensor structure of the $C_{3v}$ group is formally identical to that of $D_{3h}$, where vertical mirror reflections $\sigma_v$ enforce $\chi^{(2)}_{xxx}=0$, the co-polarized in-plane pattern is identical to the monolayer baseline:
\begin{equation}
    I_\parallel(\theta) \propto |\chi^{(2)}_{yyy}|^2\sin^2(3\theta).
\end{equation}
As a consequence, all nodes remain along the zigzag directions. The activated out-of-plane components ($\chi^{(2)}_{zzz}$, $\chi^{(2)}_{zxx}$, $\chi^{(2)}_{xxz}$) do not contribute directly to the normal-incidence co-polarized profile, although the become accessible in oblique-incidence or cross-polarized measurement settings.

\textit{$C_3$ (twisted bilayers).} In twisted configurations, both $\chi^{(2)}_{xxx}$ and 
$\chi^{(2)}_{yyy}$ are simultaneously active and independent (Eq.~\ref{eq:chi2_C3}). Evaluating the projection equations with the full $C_3$ constraints ($\chi^{(2)}_{xxy} = -\chi^{(2)}_{yyy}$, $\chi^{(2)}_{xyy} = -\chi^{(2)}_{xxx}$, $\chi^{(2)}_{yxx} = -\chi^{(2)}_{yyy}$, and $\chi^{(2)}_{yxy} = -\chi^{(2)}_{xxx}$) yields the following polarization components:
\begin{align}
    P_x &\propto \chi^{(2)}_{xxx}(\cos^2\theta - \sin^2\theta) - 2\chi^{(2)}_{yyy}\cos\theta\sin\theta, \\
    P_y &\propto \chi^{(2)}_{yyy}(\sin^2\theta - \cos^2\theta) - 2\chi^{(2)}_{xxx}\cos\theta\sin\theta.
\end{align}
Contracting these expressions into the parallel projection ($P_\parallel = P_x\cos\theta + P_y\sin\theta$) simplifies via standard trigonometric identities to:
\begin{equation}
    P_\parallel \propto -\chi^{(2)}_{yyy}\sin(3\theta) + \chi^{(2)}_{xxx}\cos(3\theta),
    \label{eq:C3_Ppara}
\end{equation}
which leads to the macroscopic co-polarized SHG intensity:
\begin{equation}
    I_\parallel(\theta) \propto \left|\chi^{(2)}_{yyy}\sin(3\theta) - \chi^{(2)}_{xxx}\cos(3\theta)\right|^2.
    \label{eq:C3_pattern}
\end{equation}
This relation can be rewritten in a compact, phase-shifted form as:
\begin{equation}
    I_\parallel(\theta) \propto \left(|\chi^{(2)}_{yyy}|^2 + |\chi^{(2)}_{xxx}|^2\right) \sin^2\left(3\left(\theta - \varphi_0\right)\right),
    \label{eq:lobeshift_pattern}
\end{equation}
where the phase-induced azimuthal lobe shift angle $\varphi_0$ is analytically defined by:
\begin{equation}
    \varphi_0 = \frac{1}{3}\arctan\!\left(\frac{\chi^{(2)}_{xxx}}{\chi^{(2)}_{yyy}}\right).
    \label{eq:phi0}
\end{equation}

Crucially, while the characteristic sixfold envelope due to threefold rotational symmetry is preserved, the entire polar lobe structure is subject to a rigid rotation by an angle $\varphi_0$ relative to the high-symmetry reference $D_{3h}$. 
Since $\chi^{(2)}_{xxx}(\omega)$ and $\chi^{(2)}_{yyy}(\omega)$ are frequency-dependent complex quantities, $\varphi_0(\omega)$ could, in principle, vary dynamically across the optical range. Whether this parameter exhibits substantial spectral variation or remains locked to the structural twist angle $\theta_{\rm tw}$ is a key question that we will address from first principles in Sec.~\ref{sec:results}.

\subsection{Accessing out-of-plane components}

The out-of-plane tensor elements $\chi^{(2)}_{zzz}$, $\chi^{(2)}_{zxx}$, and $\chi^{(2)}_{xxz}$, which are activated in the AB(3R) bilayer ($C_{3v}$) and twisted configuration ($C_3$) due to the $\sigma_h$ symmetry breaking, remain inaccessible in normal-incidence experiments where the fundamental driving field lies entirely within the $xy$ plane ($E_z = 0$). An oblique-incidence experimental setting provides a direct and elegant route to probe these otherwise inaccessible components.

When a $p$-polarized fundamental beam hits the sample at angle of incidence $\beta$ relative to the surface normal ($z$ axis), propagating in the $xz$ plane, the electric field vector possesses both in-plane and out-of-plane components:
\begin{equation}
    \mathbf{E}(\omega) = E(\omega)    \left(\cos\beta\,\hat{x} -\sin\beta\,\hat{z}\right),
    \label{eq:Epol}
\end{equation}
with $E_x = E\cos\beta$ and $E_z = -E\sin\beta$. Contracting these field components with the symmetry-allowed elements for the $C_{3v}$ and $C_3$ point groups, the second-harmonic polarization becomes:
\begin{align}
    P_z(2\omega) &\propto \chi_{zxx}\cos^2\beta +     \chi_{zzz}\sin^2\beta,
    \label{eq:Pz}\\
    P_x(2\omega) &\propto \chi_{xxx}\cos^2\beta -     \chi_{xxz}\cos\beta\sin\beta,
    \label{eq:Px}
\end{align}
where the $C_3$ rotational constraints $\chi^{(2)}_{xzz} = 0$ and $\chi^{(2)}_{zxz} = 0$ have been applied. 

As explicitly revealed by Eqs.~\eqref{eq:Px} and \eqref{eq:Pz}, the out-of-plane tensor contributions scale as $\sin^2\beta$ and $\sin\beta\cos\beta$, and vanish identically at normal incidence ($\beta = 0$) while scaling up with increasing tilt angles. An $s$-polarized beam at oblique incidence maintains $E_z = 0$, failing to activate these out-of-plane signals. Hence, $p$-polarized light is indispensable for mapping these components.

 In high-symmetry $D_{3h}$ systems, such as the isolated monolayer and the $\text{AA}$ bilayer, the horizontal reflection plane $\sigma_h$ enforces $\chi_{zzz} = \chi_{zxx} = \chi_{xxz} = 0$ regardless of $\beta$. Consequently, the probed SHG response remains invariant to changes in the incidence tilt angle $\beta$ , marking normal-incidence angular dependence. By contrast, in $C_{3v}$ and $C_3$ systems, these elements are active and introduce prominent $\beta$-dependent SHG signals.  This distinguishing behavior allows oblique-incidence measurements to provide a direct optical fingerprint of broken $\sigma_h$ symmetry, cleanly separating AB(3R) and twisted bilayers from monolayers and AA-stacks without requiring additional structural characterization.

The above-mentioned procedure can be complemented by cross-polarized detection, positioning the analyzer perpendicular to the fundamental polarization plane ($p\text{-in}/s\text{-out}$ or $s\text{-in}/p\text{-out}$). Such a setting filters out the dominant symmetric paths and isolates the $y$-component of the induced second-harmonic polarization, allowing the cross-polarized response to selectively give access to the $\chi^{(2)}_{xxz} = \chi^{(2)}_{yyz}$ elements in both $C_{3v}$ and $C_3$ symmetries, directly mapping the pure chiral tensor element $\chi^{(2)}_{xyz}$ in twisted $C_3$ superlattices. This geometry complements the co-polarized oblique-incidence measurements, forming a robust protocol for isolating the complete out-of-plane crystal response.

\section{Computational Methods}
\label{sec:bench}

%\paragraph{Density Functional Theory Calculations}
First-principles calculations were performed using the Vienna Ab initio Simulation Package (\textsc{vasp})~\cite{Vasp-1,Vasp-2} within the Perdew-Burke-Ernzerhof (PBE)~\cite{pbe-1,pbe-2} generalized gradient approximation, using the projector augmented wave method~\cite{paw} to describe core-valence interactions. Long-range van der Waals (vdW) forces were treated via the DFT-D3 correction~\cite{grimme2010dftd3}, and spin-orbit coupling was included self-consistently throughout. All considered MoS$_2$ bilayers were fully relaxed until residual interatomic forces fell below $0.01$~eV/\AA. A vacuum layer of 20~\AA{} was inserted along the out-of-plane direction to prevent spurious interactions between periodic images. Brillouin-zone sampling was performed using a $\Gamma$-centered $12\times12\times1$ $k$-mesh for all self-consistent calculations in the primitive cells, and a $6\times6\times1$ $k$-mesh for the twisted bilayer supercell.

%\paragraph{Wannier Functions and $\chi^{(2)}$ Calculation}

To compute the nonlinear optical response, maximally localized Wannier functions were constructed using \textsc{Wannier90}~\cite{wannier90v1} interfaced with \textsc{vasp}. The initial projections were chosen on the Mo $d_{z^2}$, $d_{x^2-y^2}$, $d_{xy}$, $d_{xz}$, $d_{yz}$ and S $p_x$, $p_y$, $p_z$ atomic orbitals. The Wannier fitting was performed on the $24\times24\times1$ DFT $k$-mesh for primitive cells and a $6\times6\times1$ grid for twisted supercell. The accuracy of the resulting Wannier interpolation was validated by checking that the interpolated band structures match those explicitly computed with DFT (see Supplemental Material).
These DFT calculations deliver the relaxed structure and underlying ground-state band structure. 

The frequency-dependent susceptibility $\chi^{(2)}_{ijk}(\omega)$ is subsequently evaluated from a separate first-principles response-function calculation, built on Wannier-interpolated momentum matrix elements within the independent-particle approximation. Speficically,
the frequency-dependent second-order susceptibility tensor  $\chi^{(2)}_{ijk}(\omega)$ was computed using the \textsc{postw90} utility~\cite{garcia2023,sipe2000}, which evaluates the SHG response in the velocity gauge using Wannier-interpolated momentum matrix elements. 
The total macroscopic response is decomposed into interband and intraband contributions:
\begin{equation}
    \chi^{(2)}_{ijk} = \chi^{(2),\mathrm{inter}}_{ijk} + 
    \chi^{(2),\mathrm{intra}}_{ijk},
    \label{eq:chi2_total}
\end{equation}
where the interband term involves three-band transitions:
\begin{align}
    \chi^{(2),\mathrm{inter}}_{ijk} &= \frac{e^3}{\hbar^2\omega^2} \sum_{\mathbf{k}}\sum_{n\neq m\neq l} \left[ \frac{v^i_{nm}v^j_{ml}v^k_{ln}\,f_{ln}} {(\omega_{mn}-2\omega-i\eta)(\omega_{ln}-\omega-i\eta)} \right.\nonumber\\
    &\left.\quad+ \frac{v^i_{nm}v^j_{ml}v^k_{ln}\,f_{ml}}  {(\omega_{mn}-2\omega-i\eta)(\omega_{lm}+\omega+i\eta)} \right],
    \label{eq:chi2_inter}
\end{align}
and the intraband term captures Fermi-surface contributions:
\begin{align}
    \chi^{(2),\mathrm{intra}}_{ijk} &= \frac{e^3}{\hbar^2\omega^3}    \sum_{\mathbf{k}}\sum_{n\neq m} \bigg[ \frac{v^i_{nm}(v^j_{mm}-v^j_{nn})v^k_{mn}}    {\omega_{mn}-2\omega-i\eta} \nonumber\\
    &\quad \quad \quad \quad \quad\quad+\frac{(v^i_{nm}v^j_{mn})_{;k}} {\omega_{mn}-\omega-i\eta} \bigg]f_{nm}.
    \label{eq:chi2_intra}
\end{align}
Here, the $k$-covariant derivative of the velocity matrix product along the Cartesian direction $l$ is:
\begin{equation}
    (v^i_{nm}v^j_{mn})_{;k} = \partial_{k_k}(v^i_{nm}v^j_{mn}) 
    - i(A^k_{nn} - A^k_{mm})v^i_{nm}v^j_{mn},
    \label{eq:covariant}
\end{equation}
where $A^k_{nn} = i\langle n\mathbf{k}|\partial_{k_k}|n\mathbf{k}
\rangle$ is the intra-band Berry connection of state $n$. 
Here, the notation $(\cdot)_{;k}$ denotes the $k$-covariant derivative introduced by 
Sipe and Shkrebtii~\cite{sipe2000}, in which the ordinary partial derivative $\partial_{k_k}$ of the velocity-matrix-element product is corrected by the difference of the intraband Berry connections of the two bands involved [Eq.~\eqref{eq:covariant}].
The quantities $v^i_{nm} = \langle n\mathbf{k}|\hat{v}_i|m\mathbf{k}\rangle$ represent velocity matrix elements, $\omega_{mn} = (E_{m\mathbf{k}}-E_{n\mathbf{k}})/\hbar$ denote interband transition frequencies, and $f_{nm} = f_n - f_m$ indicates the corresponding Fermi occupation differences. 

As formulated, the interband term in Eq.~\eqref{eq:chi2_inter} captures coherent virtual loops among three distinct bands ($n\to m\to l\to n$), which undergo resonant enhancements whenever the fundamental ($\hbar\omega$) or second-harmonic ($2\hbar\omega$) energies match an electronic energy gap. Concurrently, the Berry connection in Eq.~\eqref{eq:covariant} encodes the non-trivial quantum geometric phase contribution to the nonlinear polarization~\cite{garcia2023}.
The point-group selection rules derived via Neumann's principle are embedded within these products of matrix elements: for any symmetry-forbidden tensor element, individual contributions across the full Brillouin zone sum undergo pairwise cancellation.

To ensure high spectral resolution and numerical convergence of these resonant profiles, Brillouin zone integration was performed on an ultra-dense $108\times108\times1$ $k$-mesh (additional details are reported in the Supplemental Material). A Lorentzian broadening parameter $\eta = 0.01$~eV was applied across all spectra. We note that these calculations are conducted within the independent-particle approximation, in which many-body corrections are omitted. This approach aligns with the main scope of this work, aimed at establishing universal tensor structures, spatial selection rules, and geometric phase invariants, rather than reproducing experimental peak energies and intensities.

\section{Results and Analysis}
\label{sec:results}
\subsection{Validation of Tensor Structure: High-Symmetry Stackings and Twisted Bilayers}

To systematically test the symmetry-derived design rules discussed in Sec.~\ref{sec:level2}, we investigate from first principles one representative MoS$_2$ configuration for each considered point group: the freestanding 1H monolayer ($D_{3h}$), the centrosymmetric AB(2H) bilayer ($D_{3d}$), the non-centrosymmetric and AB(3R) bilayers ($C_{3v}$), and a twisted bilayer ($C_3$) with angle $\theta_{\rm tw} = 21.8^\circ$. This analysis allows us to track how the spectral weight of the frequency-dependent macroscopic $\chi^{(2)}(\omega)$ tensor shifts suppresses, or splits as the spatial symmetry elements of the parent monolayer are sequentially eliminated by stacking modifications and interlayer twisting.

\begin{figure*}%[t]
\includegraphics[width=\textwidth]{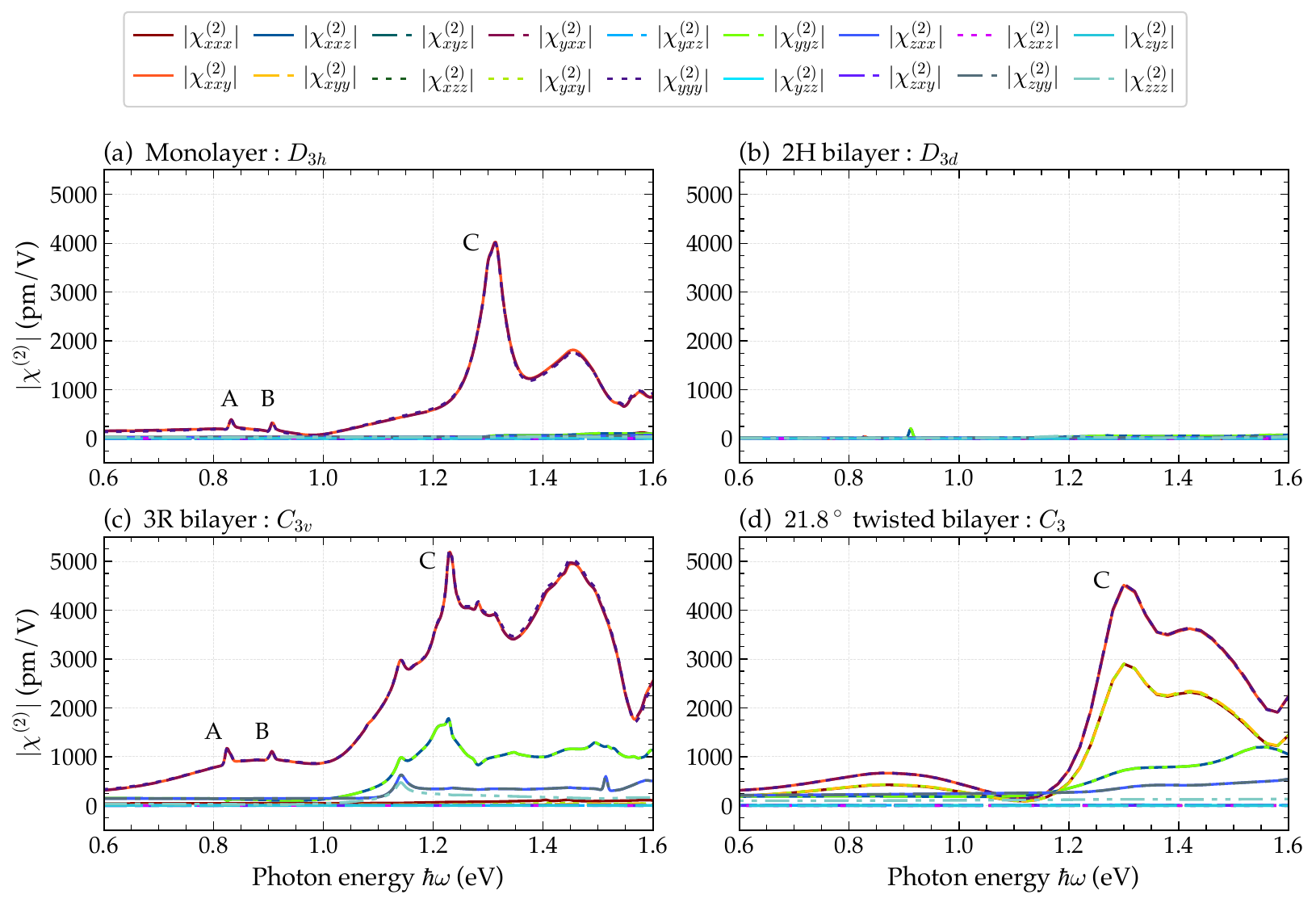}
\caption{Frequency-dependent susceptibility spectra $|\chi^{(2)}_{ijk}(\omega)|$ computed for a representative set of MoS$_2$ structures spanning different point groups with progressively reduced symmetry: $D_{3h} \to D_{3d} \to C_{3v} \to C_3$. Spectral features labeled A, B, and C correspond to key two-photon resonances.
(a) In the monolayer ($D_{3h}$), only $|\chi^{(2)}_{yyy}| \neq 0$, while all other components vanish identically by symmetry. 
(b) In the AB(2H) bilayer ($D_{3d}$), all SHG components are suppressed by the inversion center. 
(c) In the AB(3R) bilayer ($C_{3v}$), the in-plane$|\chi^{(2)}_{yyy}|$ element is retained, and the out-of-plane components $|\chi^{(2)}_{zzz}|$, $|\chi^{(2)}_{zxx}|$, and $|\chi^{(2)}_{xxz}|$ are activated by the broken horizontal reflection $\sigma_h$; the in-plane element $\chi^{(2)}_{xxx}$ remains zero under $\sigma_v$. 
(d) In the $\theta_{\rm tw} = 21.8^\circ$ twisted bilayer ($C_3$), both $|\chi^{(2)}_{yyy}|$ and $|\chi^{(2)}_{xxx}|$ emerge as independent, non-zero tensor components along with the out-of-plane contributions, confirming the unconstrained $C_3$ tensor structure.}
\label{fig:chi2_highsym}
\end{figure*}

The spectrum of the 1H monolayer exhibits only a single non-vanishing component, $|\chi^{(2)}_{yyy}|$, while all other tensor elements vanish within numerical precision [Fig.~\ref{fig:chi2_highsym}(a)]. This SHG profile matches the single-component landscape imposed by $D_{3h}$ symmetry, with the specific permutation relations 
$\chi^{(2)}_{yxx} = \chi^{(2)}_{xxy} = \chi^{(2)}_{xyx} = -\chi^{(2)}_{yyy}$ satisfied across the entire frequency range explored. 

The inclusion of a second layer completely alters the nonlinear response of MoS$_2$. In the AB(2H) bilayer stacking [Fig.~\ref{fig:chi2_highsym}(b)], all computed $\chi^{(2)}$ components vanish. This provides direct numerical verification of our structural design framework: the activation of an inversion center within the $D_{3d}$ point group completely erases the electric-dipole SHG contribution, causing destructive interlayer interference that overrides the strong nonlinear responses of the individual constituent monolayers. A detailed inspection of Fig.~\ref{fig:chi2_highsym}(b), reveals an exceptionally weak residual signal near 0.9~eV, where the monolayer hosts the B-resonance. We note that this feature does not stem from a physical violation of the crystal point group during structural relaxation, but likely originates from the Wannier function interpolation scheme, where tiny residual errors in the momentum matrix element cancellations can become visible in the vicinity of intense, localized electronic resonances. 

By rigidly translating the upper layer into the non-centrosymmetric AB(3R) stacking sequence [Fig.~\ref{fig:chi2_highsym}(c)], the inversion center is broken, and the nonlinear response is reactivated. While the in-plane component $|\chi^{(2)}_{yyy}|$ mirrors the behavior of the monolayer, the absence of the horizontal mirror plane ($\sigma_h$) activates a manifold of out-of-plane components: $|\chi^{(2)}_{zzz}|$, $|\chi^{(2)}_{zxx}|$, and $|\chi^{(2)}_{xxz}|$. Concomitantly, the retained vertical mirror planes $\sigma_v$ maintain the other in-plane component $|\chi^{(2)}_{xxx}| = 0$. 
These internal $C_{3v}$ symmetry constraints ($\chi^{(2)}_{zxx} = \chi^{(2)}_{zyy}$ and $\chi^{(2)}_{xxz} = 
\chi^{(2)}_{yyz}$) are numerically validated by our DFT results. Notably, these allowed out-of-plane elements are comparable in magnitude to the in-plane baseline $|\chi^{(2)}_{yyy}|$, indicating that breaking the horizontal reflection operation has a massive quantitative impact on the macroscopic nonlinear polarization.

Finally, eliminating all remaining mirror symmetries via relative layer twist ($C_3$ symmetry) unlocks the most complex tensor landscape, as shown in Fig.~\ref{fig:chi2_highsym}(d) for the $\theta_{\rm tw} = 21.8^\circ$ twisted bilayer. The optical fingerprint of the structural twist is manifested by the simultaneous activation of both $|\chi^{(2)}_{yyy}|$ and $|\chi^{(2)}_{xxx}|$ as independent, nonzero components. Since $\chi^{(2)}_{xxx} =0$ is no longer enforced by vertical reflection planes, its prominent spectral signature represents a direct indicator of the low-symmetry $C_3$ phase. Our numerical results faithfully reproduce the expected $C_3$ rotational invariants, confirming the equivalence $\chi^{(2)}_{xyy} = \chi^{(2)}_{yxy} = \chi^{(2)}_{yyx} = -\chi^{(2)}_{xxx}$ and $\chi^{(2)}_{yxx} = \chi^{(2)}_{xxy} = \chi^{(2)}_{xyx} = -\chi^{(2)}_{yyy}$, while verifying the activation of the out-of-plane components $|\chi^{(2)}_{zzz}|$, $|\chi^{(2)}_{zyy}|$, and $|\chi^{(2)}_{xxz}|$. 

Notably, our first-principles calculations reveal that its magnitude remains exceptionally small across the entire frequency range, even though group-theory arguments predict that the chiral out-of-plane component $\chi^{(2)}_{xyz}$ is symmetrically allowed under the $C_3$ point group. This suppression stems directly from the weak nature of the interlayer interactions. Since the electronic states remain predominantly localized within the individual monolayers, where the horizontal mirror symmetry operation is locally preserved, the global breaking of $\sigma_h$ and the resulting chiral coupling between orbitals with out-of-plane ($z$) and in-plane ($x,y$) distribution contribute only marginally to the macroscopic nonlinear susceptibility. Consequently, the in-plane elements $\chi^{(2)}_{yyy}$ and $\chi^{(2)}_{xxx}$ remain the dominant observables for detecting and characterizing twisted phases.

Finally, we note that the monolayer and the AA-stacked bilayer ($D_{3h}$ point group) possess twofold $C_2'$ rotation axes, i.e., in-plane axes oriented perpendicular to the principal $C_3$ axis~\cite{kormanyos2013}. Although these $C_2'$ operations have an impact on the nonlinear optical response, all out-of-plane $\chi^{(2)}$ components are already forced to zero by the horizontal mirror $\sigma_h$. In an idealized macroscopic twisted homobilayer, the presence of these twofold $C_2'$ axes would similarly enforce $\chi^{(2)}_{zzz} = \chi^{(2)}_{zxx} = \chi^{(2)}_{xxz} = 0$. In our DFT calculations, however, the minimal periodic supercell chosen to represent the twisted bilayer explicitly lacks this $C_2'$ axis, projecting the system strictly into the $C_3$ point group. As a consequence, the out-of-plane elements $\chi^{(2)}_{zxx}$ and $\chi^{(2)}_{xxz}$ are symmetry-allowed and yield non-zero values in our calculations.

\subsection{Two-Photon Resonances in Monolayer MoS$_2$}

\begin{figure}%[t]
\includegraphics[width=\columnwidth]{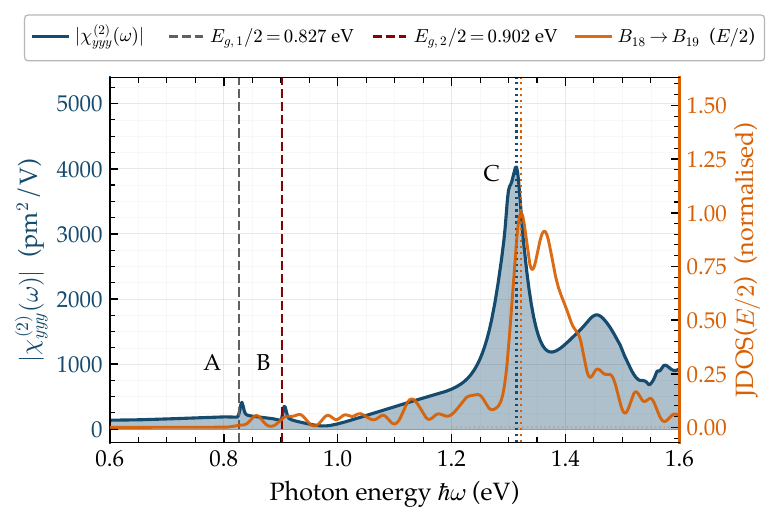}
\caption{Computed $|\chi^{(2)}_{yyy}(\omega)|$ spectrum (blue shaded area) and partial JDOS$_{\text{VBM} \to \text{CBM}}(E/2)$ (orange curve) of monolayer MoS$_2$, evaluated between the frontier valence and conduction bands ($B_{18} \to B_{19}$). Vertical dotted lines mark the A and B direct-gap transition thresholds at the K point, split by spin-orbit coupling: $E_{g,1}/2 = 0.827$~eV (A transition, VBM $\to$ CBM) and $E_{g,2}/2 = 0.902$~eV (B transition, VBM$-1$ $\to$ CBM). Two additional vertical dotted lines mark the C peak in the SHG spectrum (blue) and the aligned JDOS peak (orange), offset by $\Delta\omega = 8$~meV, confirming two-photon resonances between the VBM and CBM as the primary origin of the SHG spectral structure at the independent-particle level.}
\label{fig:jdos}
\end{figure}

To connect the spectral features of $\chi^{(2)}$ computed from first principles to the underlying electronic structure, we contrast the dominant $|\chi^{(2)}_{yyy}(\omega)|$ spectrum of monolayer MoS$_2$ against the partial joint density of states evaluated at half the photon energy, JDOS(E/2), tracking the frontier electronic manifolds (Fig.~\ref{fig:jdos}). Within the independent-particle approximation, SHG satisfies a two-photon resonance condition whenever the second-harmonic energy $2\hbar\omega$ coincides with a real, single-particle interband electronic transition. Hence, peaks appearing in the $|\chi^{(2)}_{yyy}(\omega)|$ spectrum at the fundamental photon energy $\hbar\omega$ correspond directly to features in the JDOS(E/2) plotted on the same energy axis. 

Monolayer MoS$_2$ is characterized by direct band-gap transitions at the K/K' valleys of its hexagonal Brillouin zone~\cite{mak2010,splendiani2010,xiao2012}. The valence band maximum (VBM) is split into two distinct subbands by the strong spin-orbit coupling inherent to the Mo $4d$ valence manifold. Within our single-particle band structure, the lowest-energy transition (labeled A) originates from the upper VBM to the conduction band minimum (CBM), corresponding to a single-particle two-photon threshold of $2\hbar\omega_A = 1.654$~eV ($E_{g,1}/2 = 0.827$~eV). The higher-energy feature (labeled B) stems from transitions originating at the spin-orbit-split lower valence band (VBM$-1$) targeting the CBM at $2\hbar\omega_B = 1.804$~eV ($E_{g,2}/2 = 0.902$~eV). These single-particle valley resonances are marked by vertical dashed indicators in Fig.~\ref{fig:jdos}. The close correspondence between the onset of $|\chi^{(2)}_{yyy}(\omega)|$ and these key JDOS threshold features confirms that the low-energy non-linear response is dominated by two-photon K-valley band-edge transitions.

At higher excitation energies, the $|\chi^{(2)}_{yyy}(\omega)|$ spectrum exhibits another prominent feature, labeled C, that aligns remarkably well (within $8$~meV) with the dominant global maximum of the JDOS. This correspondence confirms that the C peak originates from strong two-photon, band-nesting transitions linking parallel valence and conduction bands along the $\Gamma-K$ or $K-M$ paths within the Brillouin zone. The residual offset of $8$~meV between the $\chi^{(2)}$ maximum and the JDOS peak is consistent with the finite Lorentzian broadening ($\eta = 0.01$~eV) applied to visualize the spectra. We emphasize that within the independent-particle approximation, A, B, and C label the non-interacting single-particle transitions. While electron-hole (excitonic) binding effects will renormalize absolute peak positions downward in experimental spectra, these single-particle resonances dictate the fundamental orbital selection rules and underlying band symmetries governing the SHG response.

\subsection{SHG Polar Patterns}

\begin{figure*}%[t]
\includegraphics[width=0.85\textwidth]{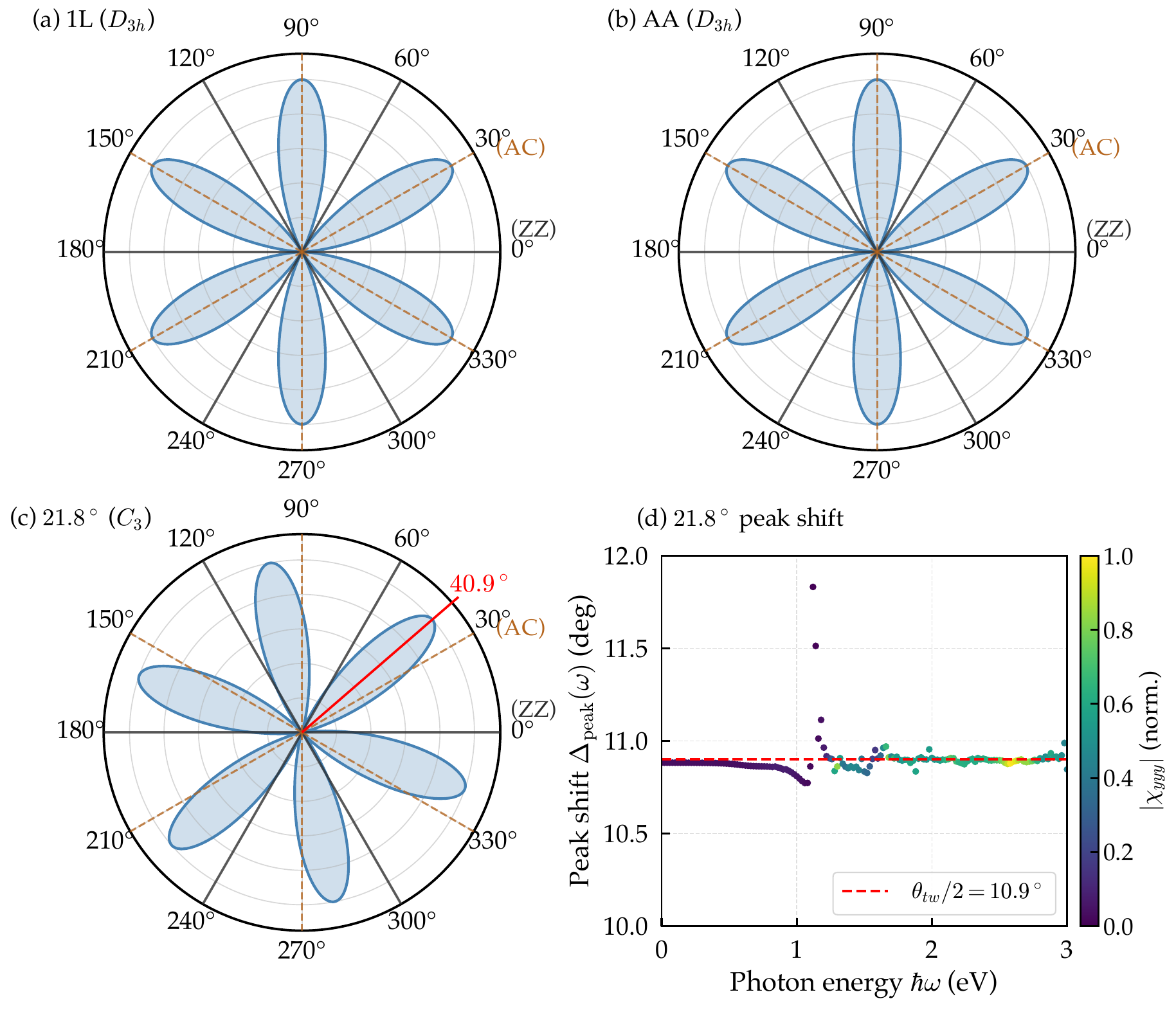}
\caption{Co-polarized SHG polar patterns $I_\parallel(\theta)$ computed for (a)~monolayer MoS$_2$ at $\hbar\omega = 1.31$~eV, (b)~AA bilayer at $\hbar\omega = 1.28$~eV, and (c)~$21.8^\circ$ twisted bilayer at $\hbar\omega = 1.30$~eV, each at their respective C-peak energy. The graphs in each panel are normalized independently to their maximum value. Orange dotted lines mark the armchair (AC)  directions. Gray solid lines mark the zigzag (ZZ) directions with the armchair (AC) axis offset by $30^\circ$. The red line in (c) marks the rotated lobe maximum at $\theta_{\rm lobe} = 40.9^\circ$, shifted by $\varphi_0 = 10.9^\circ$ due to the activation of the $\chi^{(2)}_{xxx}$ element in the $C_3$ phase. (d)~Computed SHG lobe shift angle $\varphi_0(\omega)$ as a function of photon energy $\hbar\omega$ for the $\theta_{\rm tw} = 21.8^\circ$ twisted bilayer ($C_3$). Each data point is color-coded according to the normalized amplitude $|\chi^{(2)}_{yyy}(\omega)|$. The red horizontal dashed line marks the analytical prediction $\varphi_0 = \theta_{\rm tw}/2 = 10.9^\circ$.}
\label{fig:polar}
\end{figure*}

To monitor how the stacking-controlled tensor modifications manifest in experiments, we map the angular dependence of the co-polarized SHG intensity as a function of the incident polarization angle. To connect this analysis with the electronic structure of each phase, the angular profiles are evaluated at their respective $C$-peak resonance energies: $\hbar\omega = 1.31$~eV for the monolayer, $\hbar\omega = 1.28$~eV for the AA-stacked bilayer, and $\hbar\omega = 1.30$~eV for the $\theta_{\rm tw} = 21.8^\circ$ twisted bilayer (Fig.~\ref{fig:polar}).

As shown in Figs.~\ref{fig:polar}(a) and (b), both the $1\text{H}$ monolayer and the AA-stacked bilayer display the characteristic six-lobed $\sin^2(3\theta)$ pattern predicted for the $D_{3h}$ point group (Eq.~\eqref{eq:D3h_pattern}). The nodes fall precisely along the zigzag directions ($\theta = 0^\circ, 60^\circ, 120^\circ, \dots$), whereas the intensity maxima align with the armchair directions ($\theta = 30^\circ, 90^\circ, 150^\circ, \dots$). The fact that the nodal structure and lobe orientation are identical for both configurations confirms that the macroscopic SHG polar pattern is determined entirely by the point group symmetry and remains insensitive to the number of layers as long as the spatial operations of the parent layer are preserved by the stacking arrangement.

In contrast, the $\theta_{\rm tw} = 21.8^\circ$ twisted bilayer displays a qualitatively different angular profile [Fig.~\ref{fig:polar}(c)]. While the six-lobed pattern is preserved, consistent with the retained threefold rotational symmetry ($C_3$), the entire lobe structure is rigidly rotated relative to the $D_{3h}$-symmetric reference. The principal intensity maximum, which falls at $\theta = 30^\circ$ (armchair direction) in the $D_{3h}$ phase, rotates to $\theta_{\rm lobe} = 40.86^\circ \approx 40.9^\circ$ under a $\theta_{\rm tw} = 21.8^\circ$, corresponding to a net azimuthal lobe shift of $\varphi_0 = 10.9^\circ$, in agreement with our analytical prediction [Eq.~\eqref{eq:phi0}] when evaluated from the first-principles complex tensor components $\chi^{(2)}_{xxx}$ and $\chi^{(2)}_{yyy}$ at this specific excitation energy.

To clarify whether the $C_3$ tensor structure permits a dynamically evolving SHG angular response, we evaluate the energy dependence of the phase-induced lobe shift angle $\varphi_0(\omega)$ [Fig.~\ref{fig:polar}(d)]. Since the independent tensor components $\chi^{(2)}_{xxx}(\omega)$ and $\chi^{(2)}_{yyy}(\omega)$ stem from distinct electronic transitions and exhibit different dynamical profiles, their trigonometric ratio could, in principle, undergo severe fluctuations near sharp valley resonances. 
 Activated by the broken mirror symmetries of the $C_3$ twisted stacking (Sec.~\ref{sec:design}), the calculated lobe shift remains remarkably constant around $10.9^\circ$ across the entire computed spectral range, exhibiting only minor deviations. Notably, $\varphi_0 = 10.9^\circ$ corresponds precisely to half of the macroscopic twist angle $\theta_{\rm tw} = 21.8^\circ$, demonstrating that the relative phase and amplitude ratio of the in-plane $\chi^{(2)}$ components are locked to the geometry of the twisted lattice. Within the adopted independent-particle approximation, where dynamical correlations are neglected, the relative ratio $\chi^{(2)}_{xxx}/\chi^{(2)}_{yyy}$ is solely dictated by the spatial projection of the atomic coordinates rather than evolving independently with the individual interband transitions. An analytical proof demonstrating how coherent field superposition between rigidly rotated monolayers locks the tensor ratio to $\tan(3\theta_{\rm tw}/2)$, and consequently forces $\varphi_0 = \theta_{\rm tw}/2$ across arbitrary twist angles, is provided in Appendix~\ref{app:phi0}.

This geometric interpretation is further substantiated by the color-coded spectral weight mapping in Fig.~\ref{fig:polar} (d). In energy windows where the nonlinear signal is robust, matching the A, B, and C two-photon resonances, the data points converge onto the ideal $\theta_{\rm tw}/2 = 10.90^\circ$ baseline. Minor fluctuations occur exclusively where the magnitude of the tensor components approaches zero, rendering the trigonometric ratio in the definition of $\varphi_0$ (Eq.~\ref{eq:phi0}) highly sensitive to numerical background noise.

\section{Conclusions}
In summary, we have presented a unified analytical and first-principles investigation of SHG in TMD subjected to different bilayer stackings, taking MoS$_2$ as the reference material and the freestanding monolayer as the structural baseline. Using Neumann's principle, we derived the complete $\chi^{(2)}$ tensor structure for each considered point group ($D_{3h}$, $D_{3d}$, $C_{3v}$, and $C_3$) and established a set of design rules connecting stacking configuration to nonlinear optical response.

Our structural framework resolves into the following key physical principles:
\begin{itemize}
    \item \textbf{Inversion Symmetry and SHG Elimination:} The activation of a macroscopic inversion center within the $\text{AB}(2\text{H})$ configuration ($D_{3d}$ point group) enforces a strict $\chi^{(2)} = 0$ identity, overriding the strong nonlinear response of the constituent layers via destructive interlayer interference.
    \item \textbf{Horizontal Mirror Plane and Out-of-Plane Paths:} The horizontal mirror plane $\sigma_h$ inherent to the monolayer and the $\text{AA}$ bilayer ($D_{3h}$) strictly forbids all out-of-plane $\chi^{(2)}$ tensor elements. Conversely, removing $\sigma_h$ in $\text{AB}(3\text{R})$ stackings ($C_{3v}$) and twisted bilayers ($C_3$) activates the $\chi^{(2)}_{zzz}$, $\chi^{(2)}_{zxx}$, and $\chi^{(2)}_{xxz}$ elements, creating an experimental path to probe hidden vertical polarization vectors via oblique-incidence $p$-polarized geometries.
    \item \textbf{Vertical Mirror Plane and Lobe Rotation:} The vertical mirror plane $\sigma_v$ acts as the defining operation that forces $\chi^{(2)}_{xxx} = 0$ in all non-twisted bilayers. Releasing this constraint through artificial twisting ($C_3$) unlocks $\chi^{(2)}_{xxx}$ as a fully independent, active tensor component, driving a rigid spatial rotation of the macroscopic co-polarized SHG polar lobes.
\end{itemize}

Finally, our first-principles calculations reveal that this phase-induced azimuthal lobe shift angle $\varphi_0$ is extraordinarily stable across the explored optical window. It remains locked to the clear geometric limit dictated by the spatial projection of the atomic coordinates:
\begin{equation}
    \theta_{tw} = 2\varphi_0.
\end{equation}
This relation follows directly from the coherent superposition of two rigidly rotated, identical monolayer responses (Appendix~\ref{app:phi0}), and holds for arbitrary twist angle in this weak-coupling regime, not being specific to $\theta_{tw}=21.8^\circ$. As such, it provides an elegant, material-agnostic, and non-destructive optical protocol for mapping localized twist profiles and structural moiré domains, at twist angles for which the twisted bilayer retains its rigid, uniformly-rotated character.

The design principles formulated in this work are universal, being dictated by global crystal symmetries rather than material-specific details. As such, their application is straightforward to all 2H-TMD bilayers and other vdW crystals with analogous structural symmetries. Most importantly, the proposed analytical framework provides a symmetry-based roadmap for engineering the nonlinear optical response of layered materials solely through stacking control. This complements parallel advances in symmetry reduction via Janus engineering and in-plane heterostructuring aimed at extreme directional SHG anisotropy~\cite{bao2026}, together forming a comprehensive paradigm for tailoring $2\text{D}$ nonlinear optics. Concurrently, the relation $\theta_{tw} = 2\varphi_0$ provides a simple and experimentally accessible optical probe of twist angle that warrants further investigation across a broader range of commensurate and incommensurate twisted TMD bilayers, paving the way for high-throughput screening of this novel material class with a non-invasive optical tool.

\section*{Acknowledgement}
The authors thank Giancarlo Soavi and Michele Guerrini for stimulating discussions. This work was funded by the German Research Foundation, project numbers  398816777 (CRC 1375 ``NOA'', subproject A8) and 547611111 (WHAT-A-TWIST). Computational resources were provided by the German National High-Performance Computing Alliance, project ID nip00092.

\section*{Data Availability}
The data collected in this work are available free of charge on Zenodo at the following DOI: 10.5281/zenodo.21480888

\appendix
\section{Symmetry Operations on $\chi^{(2)}$}
\label{app:symmetry}
The second-order susceptibility $\chi^{(2)}_{ijk}$ is a third-rank polar tensor with $3^3 = 27$ elements. In this Appendix, we provide the explicit algebraic reduction of these components under intrinsic permutation symmetry and relevant point group operations based on Neumann's principle.

\subsection{Intrinsic Permutation Symmetry}
In the SHG experimental setting, two incoming photons oscillate at the identical frequency $\omega$, making the last two indices of $\chi^{(2)}_{ijk}$ completely interchangeable:
\begin{equation}
\chi^{(2)}_{ijk} = \chi^{(2)}_{ikj}.
    \label{eq:perm}
\end{equation}
For any fixed index $i \in \{x,y,z\}$, the remaining $jk$ pair possesses $3\times3=9$ permutations, only 6 of which are linearly independent under $j \leftrightarrow k$ exchange:
\begin{equation}
    \{jk\} \in \{xx,\, yy,\, zz,\, xy(=yx),\, xz(=zx),\, yz(=zy)\}.
\end{equation}
This constraint reduces the 27 initial components to $3 \times 6 = 18$ independent elements.

\subsection{Reduction by $C_3$ Rotation}
The threefold rotation operation $C_3$ around the out-of-plane $z$-axis represents a counterclockwise rotation by $120^\circ$ about the $z$-axis, defined by the standard transformation matrix:
\begin{equation}
    R(C_3) = \begin{pmatrix} 
    -\frac{1}{2} & -\frac{\sqrt{3}}{2} & 0 \\ 
    \frac{\sqrt{3}}{2} & -\frac{1}{2} & 0 \\ 
    0 & 0 & 1 
    \end{pmatrix}.
    \label{eq:C3matrix}
\end{equation}
Neumann's principle requires that the tensor must remain invariant under any valid symmetry operations of the crystal point group:
\begin{equation}
    \chi^{(2)}_{ijk} = R_{il}R_{jm}R_{kn}\,\chi^{(2)}_{lmn}.
    \label{eq:neumann}
\end{equation}
Since $R_{zz} = 1$ and $R_{xz} = R_{yz} = R_{zx} = R_{zy} = 0$, the $z$ axis is invariant under $C_3$. This has two main consequences:

\textit{(i) Elements mixing one in-plane index with two out-of-plane indices.} Taking $\chi^{(2)}_{xzz}$ as representative baseline and noting that $R_{zm}=\delta_{zm}$ isolates the out-of-plane projection, the transformation simplifies to $\chi^{(2)}_{xzz} = 
R_{xl}\chi^{(2)}_{lzz}$. Expanding over $l \in \{x,y,z\}$ yields:
\begin{align}
    \chi^{(2)}_{xzz} &= R_{xx}\chi^{(2)}_{xzz} + R_{xy}\chi^{(2)}_{yzz} + 
    R_{xz}\chi^{(2)}_{zzz} \nonumber\\
    &= -\tfrac{1}{2}\chi^{(2)}_{xzz} - \tfrac{\sqrt{3}}{2}\chi^{(2)}_{yzz} 
    + 0,
\end{align}
which reduces to $\tfrac{3}{2}\chi^{(2)}_{xzz} = -\tfrac{\sqrt{3}}{2}\chi^{(2)}_{yzz}$.
Performing the identical operation for the complementary component $\chi^{(2)}_{yzz}$ results in:
\begin{align}
    \chi^{(2)}_{yzz} &= R_{yx}\chi^{(2)}_{xzz} + R_{yy}\chi^{(2)}_{yzz} + 
    R_{yz}\chi^{(2)}_{zzz} \nonumber\\
    &= \tfrac{\sqrt{3}}{2}\chi^{(2)}_{xzz} - \tfrac{1}{2}\chi^{(2)}_{yzz} 
    + 0,
\end{align}
which simplifies to $\tfrac{3}{2}\chi^{(2)}_{yzz} = \tfrac{\sqrt{3}}{2}\chi^{(2)}_{xzz}$.
Solving this paired system of linear equations simultaneously forces both components to vanish:
\begin{equation}
    \chi^{(2)}_{xzz} = \chi^{(2)}_{yzz} = 0.
\end{equation}
The same algebra applied to $\chi^{(2)}_{zxz}$, $\chi^{(2)}_{zzx}$, 
$\chi^{(2)}_{zyz}$, $\chi^{(2)}_{zzy}$ gives:
\begin{equation}
    \chi^{(2)}_{zxz} = \chi^{(2)}_{zzx} = \chi^{(2)}_{zyz} = \chi^{(2)}_{zzy} = 0.
\end{equation}

\textit{(ii) Elimination of mixed in-plane elements $\chi^{(2)}_{zxy}$.} Applying Eq.~\eqref{eq:neumann} to $\chi^{(2)}_{zxy}$ requires expanding across the entire in-plane manifold ($m,n \in \{x,y\}$):
\begin{align}
    \chi^{(2)}_{zxy} &= R_{xm}R_{yn}\chi^{(2)}_{zmn} \nonumber\\
    &= -\frac{\sqrt{3}}{4}\chi^{(2)}_{zxx} + \frac{1}{4}\chi^{(2)}_{zxy} 
    - \frac{3}{4}\chi^{(2)}_{zxy} + \frac{\sqrt{3}}{4}\chi^{(2)}_{zyy}.
\end{align}
Enforcing permutation symmetry $\chi^{(2)}_{zyx} = \chi^{(2)}_{zxy}$ alongside the structural constraint $\chi^{(2)}_{zxx} = \chi^{(2)}_{zyy}$, which can be verified independently by passing $\chi^{(2)}_{zxx}$ through the identical $C_3$ transformation matrix:
\begin{align}
    \chi^{(2)}_{zxx} &= R_{xm}R_{xn}\chi^{(2)}_{zmn} \nonumber\\
    &= \tfrac{1}{4}\chi^{(2)}_{zxx} - \tfrac{\sqrt{3}}{4}\chi^{(2)}_{zyx} 
    - \tfrac{\sqrt{3}}{4}\chi^{(2)}_{zxy} + \tfrac{3}{4}\chi^{(2)}_{zyy},
\end{align}
the structural equation reduces neatly to $\frac{3}{2}\chi^{(2)}_{zxy} = 0$. Hence, $\chi^{(2)}_{zxy} = \chi^{(2)}_{zyx} = 0$. 

Overall, the $C_3$ rotation symmetry nullifies 8 tensor components:
\begin{equation}
    \chi^{(2)}_{xzz} = \chi^{(2)}_{yzz} = \chi^{(2)}_{zxz} = \chi^{(2)}_{zzx} = 
    \chi^{(2)}_{zyz} = \chi^{(2)}_{zzy} = \chi^{(2)}_{zxy} = \chi^{(2)}_{zyx} = 0.
    \label{eq:C3zeros}
\end{equation}
Taking into account permutation rules, this condition leaves 13 unique independent tensor components within the $C_3$ point group.

\subsection{Horizontal Mirror Plane $\sigma_h$}
The horizontal mirror plane $\sigma_h$ reflects spatial coordinates across the basal plane, mapping $x \to x$, $y \to y$, $z \to -z$, with matrix $R(\sigma_h) = \text{diag}(1,1,-1)$. Under this transformation, each $z$ index ($n_z$) contributes a factor of $-1$:
\begin{equation}
    \chi^{(2)}_{ijk} \xrightarrow{\sigma_h} (-1)^{n_z}\chi^{(2)}_{ijk}.
\end{equation}
Applying Neumann's principle then forces $\chi^{(2)}_{ijk} = 0$ for any combination where $n_z$ is an odd integer. 
Taking $\chi^{(2)}_{xxz}$ ($n_z=1$) as an explicit example, we obtain:
\begin{align}
    \chi^{(2)}_{xxz} & =  R_{xx}R_{xx}R_{zz}\chi^{(2)}_{xxz}  \nonumber\\
    &= (1)(1)(-1)\chi^{(2)}_{xxz} = -\chi^{(2)}_{xxz} \implies \chi^{(2)}_{xxz} = 0.
\end{align}
Consequently, all tensor components with an odd number of out-of-plane coordinates vanish completely:
\begin{equation}
    \chi^{(2)}_{xxz},\, \chi^{(2)}_{yyz},\, \chi^{(2)}_{zxx},\, \chi^{(2)}_{zyy},\, 
    \chi^{(2)}_{xyz},\, \chi^{(2)}_{xzy},\, \chi^{(2)}_{yxz},\, \chi^{(2)}_{yzx},\, 
    \chi^{(2)}_{zzz} = 0.
\end{equation}
Components containing purely in-plane or an even number of out-of-plane coordinates are unaffected.

\subsection{Vertical Mirror Plane $\sigma_v$}
The vertical mirror plane $\sigma_v$ reflects coordinates across a vertical plane perpendicular to the layer, mapping $x \to x$, $y \to -y$, $z \to z$, expressed as $R(\sigma_v) = \text{diag}(1,-1,1)$. Here, each $y$ index ($n_y$) contributes a factor of $-1$:
\begin{equation}
    \chi^{(2)}_{ijk} \xrightarrow{\sigma_v} (-1)^{n_y}\chi^{(2)}_{ijk}.
\end{equation}
Components with an odd value of $n_y$ vanish. Taking as an example the in-plane component $\chi^{(2)}_{xxx}$, the $C_3$ rotation symmetry enforces the relation $\chi^{(2)}_{xxx} = -\chi^{(2)}_{xyy}$. Subjecting these elements to $\sigma_v$ yields:
\begin{equation}
    \chi^{(2)}_{xxx} \xrightarrow{\sigma_v} \chi^{(2)}_{xxx}, \qquad
    \chi^{(2)}_{xyy} \xrightarrow{\sigma_v} -\chi^{(2)}_{xyy},
\end{equation}
Since Neumann's principle requires $\chi^{(2)}_{xyy} = -\chi^{(2)}_{xyy} = 0$, the constraint reflects back through the rotational relation, ensuring $\chi^{(2)}_{xxx} = 0$. Similarly, any component containing a $y$-index, such as $\chi^{(2)}_{xyz}$ ($n_y=1$), is systematically canceled:
\begin{equation}
    \chi^{(2)}_{xyz} \xrightarrow{\sigma_v} -\chi^{(2)}_{xyz} 
    \implies \chi^{(2)}_{xyz} = 0.
\end{equation}
This structural constraint applies to both $D_{3h}$ and $C_{3v}$ point groups. The absence of vertical mirror plane $\sigma_v$ in the $C_3$ point group is the exclusive mechanism unlocking $\chi^{(2)}_{xxx}$ and $\chi^{(2)}_{xyz}$ as active, independent tensor elements in twisted bilayers.

\subsection{Inversion Symmetry $\mathcal{I}$}
A macroscopic spatial inversion operation maps all coordinates through the origin, $\mathbf{r} \to -\mathbf{r}$, yielding a transformation matrix $R(\mathcal{I}) = \text{diag}(-1,-1,-1)$. For a third-rank polar tensor, this operation introduces an odd transformation parity across all indices simultaneously:
\begin{equation}
    \chi^{(2)}_{ijk} = 0 \qquad \forall\; i,j,k.
\end{equation}
Enforcing Neumann's principle requires $\chi^{(2)}_{ijk} = -\chi^{(2)}_{ijk}$, which in turn implies:
\begin{equation}
    \chi^{(2)}_{ijk} = 0 \qquad \forall\; i,j,k.
\end{equation}
This complete elimination of tensor elements explains why all electric-dipole SHG paths are completely prohibited within the centrosymmetric 2H bilayer structure ($D_{3d}$ point group).

\section{Analytical Derivation of the Twist-Angle Locking $\varphi_0 = \theta_{tw}/2$}
\label{app:phi0}

We analytically establish that the relation $\varphi_0 = \theta_{\mathrm{tw}}/2$ represents a universal geometric property of weakly coupled $C_3$-symmetric twisted bilayers.
In the weak-interlayer-coupling limit, a twisted bilayer behaves as two $D_{3h}$-symmetric monolayers with identical intrinsic in-plane susceptibility $\chi_0(\omega) \equiv \chi^{(2)}_{yyy}(\omega)$, rotated by a relative twist angle $\theta_{tw}$, whose generated SHG fields superpose coherently. As established in Eq.~(\ref{eq:p_parallel}), the co-polarized nonlinear polarization of the bottom layer ($\theta_{tw} = 0^\circ$) is:
\begin{equation}
P_\parallel^{(1)}(\theta) \propto -\chi_{yyy} \sin(3\theta).
\end{equation}
For the top layer, rotated by $\theta_{tw}$, the armchair direction shifts by $\theta_{tw}$, yielding:
\begin{equation}
P_\parallel^{(2)}(\theta) \propto -\chi_{yyy} \sin\!\big[3(\theta-\theta_{tw})\big].
\end{equation}
The total macroscopic co-polarized response is given by the linear superposition:
\begin{equation}
P_\parallel^{\text{tot}}(\theta) \propto -\chi_{yyy}\Big[\sin 3\theta + \sin(3\theta - 3\theta_{tw})\Big].
\label{eq:Ptot}
\end{equation}
Applying the sum-to-product trigonometric identity $\sin A + \sin B = 2\sin\!\left(\dfrac{A+B}{2}\right) \cos\left(\dfrac{A-B}{2}\right)$, Eq.~\eqref{eq:Ptot} reduces to:
\begin{equation}
P_\parallel^{\text{tot}}(\theta) \propto -2\chi_{yyy} \cos\!\left(\frac{3\theta_{tw}}{2}\right)
\sin\!\left[3\left(\theta - \frac{\theta_{tw}}{2}\right)\right].
\label{eq:phi0-derivation}
\end{equation}
Comparing Eq.~\eqref{eq:phi0-derivation} with the general $C_3$ polar pattern expression, $P_\parallel(\theta) \propto -C\sin[3(\theta-\varphi_0)]$, delivers the geometric identity:
\begin{equation}
\varphi_0 = \frac{\theta_{tw}}{2}, \qquad C = 2\chi_{yyy}\cos\!\left(\frac{3\theta_{tw}}{2}\right),
\label{eq:phi0-result}
\end{equation}
Equivalently, this coherent field superposition locks the ratio of the effective $C_3$ tensor components to $\chi^{(2)}_{xxx}/\chi^{(2)}_{yyy} = \tan(3\theta_{\mathrm{tw}}/2)$, enforcing $\varphi_0 = \theta_{\mathrm{tw}}/2$ strictly independently of photon energy $\hbar\omega$ or the absolute magnitude of $\chi_0(\omega)$.

The general relation $\varphi_0 = \frac{1}{3}\arctan\big(\chi^{(2)}_{xxx}/\chi^{(2)}_{yyy}\big)$ [Eq.~(21)] is exact and depends solely on $C_3$ point-group symmetry. However, the geometric lock $\chi^{(2)}_{xxx}/\chi^{(2)}_{yyy} = \tan(3\theta_{\mathrm{tw}}/2)$ assumes a rigid, uniform rotational superposition of intact monolayers. In the small-angle regime ($\theta_{\mathrm{tw}} \lesssim 2^\circ$), atomic reconstruction induces local domain-wall formation (e.g., periodic 3R-like and AA-like domains~\cite{Xu2024}). In this strongly reconstructed limit, local interlayer sliding overrides the uniform rotational mapping, modifying the tensor component ratio from its weak-coupling geometric limit.

\bibliography{apssamp}% Produces the bibliography via BibTeX.

\end{document}